\documentclass[11pt,twoside]{pnas-new}

\templatetype{pnasresearcharticle} % Choose template 
\usepackage{comment}
\usepackage{dcolumn}% Align table columns on decimal point
\newcolumntype{d}[1]{D{.}{.}{#1}}
\usepackage{bm}% bold math
\usepackage[capitalise,nameinlink]{cleveref}
\usepackage{chemformula}
\usepackage[disable]{todonotes}
\usepackage{multirow}
\usepackage{multicol}
\usepackage{pdfpages}

\usepackage{setspace}

\graphicspath{
	{./figures-arxiv/}
}

\title{Nearsightedness of Crystalline Materials and Intergranular Embrittlement}%Crystalline Materials: Nearsighted and Cataractous%Nearsightedness and Cataracts (or as adjectives, Nearsighted and Cataractous)}

\author[a]{M.\ Rajivmoorthy}
\author[a]{T.\ R.\ Wilson}
\author[a,*]{M.\ E.\ Eberhart}

\affil[a]{Molecular Theory Group, Colorado School of Mines, Golden, Colorado, USA}
\affil[*]{To whom correspondence should be addressed. E-mail: \href{mailto:meberhar@mines.edu?subject=Nearsightedness\%20of\%20Crystalline Materials\%20and\%20Intergranular\%20Embrittlement\%20(Nature\%20Submission)}{meberhar@mines.edu}}

\leadauthor{Rajivmoorthy}


\dates{This manuscript was compiled on \today}

\doi{Preprint version (arXiv.org)}

\begin{abstract}
Our quest  to design  materials often envisions as a first step the conceptual decomposition of a material into meaningful atomic scale neighborhoods.
The performance of the monolithic material is then seen to arise from the combined properties of these much simpler regions.
It is the nearsightedness of electronic matter (NEM) principle that provides the rigorous justification for this ``divide and conquer'' approach.
NEM asserts that a material property may be significantly affected by a perturbation, no matter how large, only over a neighborhood of size $\bm{R}$.
Though NEM posits the existence of meaningful atomic scale neighborhoods, for the most part these regions are identified empirically.
In this paper we propose a  methodology to divide real materials into meaningful neighborhoods determined  by the topology of the charge density.
We generalize this approach by applying the same  to determine neighborhoods representative of elemental crystalline materials and then use these neighborhoods to model  the embrittling effects of bismuth atoms segregated to copper grain boundaries. 
We show that embrittlement is the result of impurity atom-induced enhancements of copper nearsightedness.  
We further suggest that just as nearsightedness  plays an overlooked role mediating embrittlement, it may also be an  important  factor affecting a broad range of unresolved problems  as apparently diverse as energy focussing phenomena and enzyme kinetics. 
\end{abstract}

\begin{document}

\maketitle
\thispagestyle{firststyle}

\abscontent

%\hrule
%\tableofcontents\vspace{1.2em}
%\hrule
%\listoftodos
%\hrule

\section{Introduction}
%\doublespacing

%%%%% summary paragraph %%%%%

%%%%% end summary paragraph %%%%%

%\dropcap{T}
The performance of structural materials derives from the way strain energy is distributed among the material's many defects and other heterogeneities. While elasticity theory and continuum mechanics are useful in describing the effects of defects on the long-range distribution of strain, the effects due to atomic scale fluctuations  are poorly understood.  As a quintessential example, normally ductile materials like copper and iron  become brittle in the presence of even minuscule quantities of elements such as hydrogen or sulfur.  Obviously, embrittling elements alter the distribution of strain energy, though the question of how these effects are produced  remains open.  

Embrittlement is but one technologically important problem in the broader class of phenomena controlled by atomic scale influences to the energy of an applied perturbation.  While  developing a more fundamental understanding of the mechanisms responsible for these processes would enhance our ability to select  compositions  to optimized materials properties, the  diversity of defects and heterogeneities that must be considered makes this a daunting task.  Even so, investigators are pursuing  methodologies intended to systematize this effort. One such approach appeals to the  nearsightedness of electronic matter principle (NEM), which asserts that the properties of complex materials can be considered to arise piecewise from local neighborhoods.\cite{NEM_PRL, NEM_PNAS, NEM_PRB}  In this approach,  neighborhoods may be investigated one at a time using ``divide and conquer'' techniques through which large and complex molecules and solids are decomposed (divided) into neighborhoods with comprehensible properties (conquered).\cite{Divide_conqer_anastassia}

NEM derives its name from the observation that atoms ``see clearly'' only nearby atoms.
Rigorously, it posits that for a fixed chemical potential the charge density, $\rho(\bm{r})$, and local properties originating from $\rho(\bm{r})$ are sensitive to changes to the external potential within some radius $R$.
Changes to the potential beyond $R$---no matter how large---do not significantly affect the density.\cite{NEM_PNAS}
In other words, there is a monotonically decreasing function, $\Delta \rho(\Omega_{\bm{r}_0}, R)$, giving the cumulative density change over a connected region $\Omega$  about $\bm{r}_0$ due to a perturbation of any magnitude at $R$, such that $\lim_{R \to \infty} \Delta \rho = 0$.  As a useful corollary, if $\Omega$ is a region over which the energy is well-defined, then  there is a function $\Delta \varepsilon_{\Omega}(R)$ giving the energy change of $\Omega$ due to a perturbation of any magnitude a distance $R$ from $\Omega$ such that $\lim_{R\to \infty} \Delta \varepsilon_{\Omega}= 0$.

In its energy form NEM provides the means to locate atomic neighborhoods from which a property, $p$, originates.
One begins by isolating a region  $\Omega$ of radius $R_{\Omega}$ from its larger molecule or solid. In principle one may calculate the energy of this isolated region and the change to its energy as  the  atomic environment at successively  greater distances  from $\Omega$  are restored.   At some   $R$, designated $R_p$,   $\Delta \varepsilon_{\Omega} $ becomes small relative to the property's characteristic energy. In this way, the property may be argued to emerge from the  atomic structure of a neighborhood of radius $R_{\Omega} + R_p $, where $R_p$ is ostensibly the width of the boundary  separating $\Omega$ from a free surface.

Equally consequential, the electronic structure of these local neighborhoods may be considered  essentially independent, making it possible  to use massively parallel algorithms to calculate the structure of large and complex materials.\cite{divide_conquer, Divide_conqer2, Divide_conqer3, NEM_PRL}
The inherent inaccuracy associated with such schemes depends on the magnitude of $\Delta \varepsilon_{\Omega}(R_p)$, which  NEM asserts decreases as $R_p$ increases. For such calculatons the size of neighborhoods and the widths of their boundaries are chosen so as to minimize computational time while achieving some desired energy accuracy.  In these instances, $R_p$ is treated as an adjustable parameter rather than a property in and of itself.  

Here we report our efforts to directly calculate $\Delta \varepsilon_{\Omega}(R)$ across a series of crystalline materials. We then apply these findings to identify the atomic neighborhoods mediating metallic grain boundary properties and, using the copper bismuth system as an example, show that nearsightedness plays an important role mediating the effects of impurities to a material's mechanical properties.  

\section{Calculations and Results}

%Our approach to determining neighborhood size and hence  nearsightedness begins by calculating the energy of a crystalline region as a large free surface is moved to greater distances, effectively simulating a ``growing'' crystalline cluster. Because of its conceptual and computatioal simplicity, this growing cluster model is ideally suited to the investigation of nearsightedness.

Approximately eighty clusters representing crystals of eleven elements 
%(Al, Si, V, Cu, Nb, Mo, Tc, Ru, Rh, Pd, Ag) 
were constructed from a central atom and its first $n$ coordination shells (concentric spheres containing the nearest neighbor atoms, the second nearest neighbor atoms, and so on to the $n^{\text{th}}$ nearest neighbors) with $n$ varying from 0 to as large as 11.  We defined the region $\Omega$, also called the central cluster, to be  a central atom and its first coordination shell.
 
These clusters were then modeled with DFT methods provided within the SCM chemistry and materials modeling suite (see SI for more details).\cite{ADF1, ADF3, band1, band2}  The per atom energy of  $\Omega$ was found using Bader partitioning.\cite{AIM, matta}   Through this combination of techniques we were able to calculate an effective central atom energy  as perturbed by an increasing number of coordination spheres designated $E^x_n$, where $x$ indicates the element and $n$ the number of coordinations spheres surrounding the central atom.
The perturbation energy to this effective  central  atom as a function of $n$ is then given by $E^x_n - E^x_0 \equiv \Delta E^x_n $.
A zero of energy may be established by noting that as $n$ grows without bound, $ \Delta E^x_n$ will approach the crystal's formation energy, $E^x_{\! f}$, a quantity readily computed using band methods.
Consequently, we define a nearsightedness function \mbox{$\Delta \varepsilon^x_n \equiv \,\mid E^x_{\! f} - \Delta E^x_n \mid$} and note  that $\lim_{n\to \infty} \Delta \varepsilon^x_n = 0$.  A property dependent nearsightedness distance can be extracted from this funtion as the values of $n$ for which $ \Delta \varepsilon^x_n$ becomes smaller than the characteristic energy of the property of interest.

We constructed our computations in three sets of increasing computational complexity and hence accompanied by greater imprecision in the calculated form of the nearsightedness function.  
Using these constructions,  we were able to apply what was learned from the less complex models to deduce trends to the form of $\Delta \varepsilon^x_n$  that may have eluded us  had we proceeded directly to the final calculations. 
The first set of calculations determined $\Delta \varepsilon^x_n$  of four comparatively light crystals: diamond cubic (DC) silicon, a prototype covalent material; face centered cubic (FCC) aluminum, a free electron metal; FCC copper, a  $d$-block metal with a full $d$-band; and body centered cubic (BCC) vanadium, a metal with a partially occupied $d$-band.  
%Accurate Bader atom energies were achieved with a triple zeta basis set including two polarization terms (TZ2P) basis set for all these crystals. 
 %Hence both cluster and BAND calculations used this basis set. 
The elements of the second set were drawn from the heavier 4$d$ transition metals and contain the BCC metals niobium and molybdenum; the hexagonal close packed (HCP) metals technetium and ruthenium; and the FCC metals rhodium, palladium, and silver. 
Compared to set 1, larger basis sets, necessary to model these crystals, were unavailable, introducing basis set error in the determination of the crystalline formation energy. 
%Compared to set 1, larger basis sets were need to model these crystals, which introduced basis set error in the determination of the crystalline formation energy. 
%Accurate Bader atom energies for this group required a quadruple zeta basis set including four polarization terms (QZ4P) basis set that was not available to the BAND calculations, which instead utilized a TZ2P basis set.
%The larger basis set required for this group of crystals increased computational complexity, and, compared to the first set, reduced the maximum cluster size that could be modeled. }
And for the third set we modeled a pure copper grain boundary along with the same boundary containing bismuth atoms.  
The complicating factor for this set of calculations was to assess the effect  of non-crystallinity on the nearsightedness function.

\subsection{Silicon, Aluminum, Vanadium and Copper}

% Copied from Microsoft Excel
% Workbook: /Users/haiiro/Dropbox/Nearsightedness:Tables.xlsx
% Worksheet: Sheet1
% Range: $Y$9:$AK$12
\begin{table}[h!]
	\centering
	\caption{\label{table:DC}
		The diamond cubic (DC) shell structure.
		Row 1:~Number of the coordination shell.
		Coordination shell zero is the central atom.
		Row 2:~Number of atoms in coordination shell $\bm{n}$.
		Row 3:~Total number of atoms in the cluster of $\bm{n}$ coordination shells.
		(Hard sphere representations of some of these clusters are provided in the SI.)
		Row 4:~Radius of the cluster, i.e.~distance between the central atom and the atoms of the $\bm{n^{\text{th}}}$ shell in atomic diameters or equivalently nearest neighbor separations.}
	\begin{tabular}{|l|c|c|c|c|c|c|c|c|c|c|c|c|}
		\hline
		\textbf{Coordination shell $\bm{n}$}
		& \textbf{0}
		& \textbf{1}
		& \textbf{2}
		& \textbf{3}
		& \textbf{4}
		& \textbf{5}
		& \textbf{6}
		& \textbf{7}
		& \textbf{8}
		& \textbf{9}
		& \textbf{10}
		& \textbf{11}
		\\\hline
		Number of $n^{\rm{th}}$ neighbors
		& 0
		& 4
		& 12
		& 12
		& 6
		& 12
		& 24
		& 16
		& 12
		& 24
		& 12
		& 8
		\\\hline
		Total atoms in cluster
		& 1
		& 5
		& 17
		& 29
		& 35
		& 47
		& 71
		& 87
		& 99
		& 123
		& 135
		& 143
		\\\hline
		Cluster radius
		& 0
		& 1
		& $\text{2} \sqrt{\frac{\text{2}}{\text{3}}}$
		& $\sqrt{\frac{\text{11}}{\text{3}}}$
		& $\text{4}\sqrt{\frac{\text{1}}{\text{3}}}$
		& $\sqrt{\frac{\text{19}}{\text{3}}}$
		& $\text{2} \sqrt{\text{2}}$
		& 3
		& $\text{4} \sqrt{\frac{\text{2}}{\text{3}}}$
		& $\sqrt{\frac{\text{35}}{\text{3}}}$
		& $\sqrt{\frac{\text{43}}{\text{3}}}$
		& 4
		\\
		\hline
	\end{tabular}
\end{table}

% Copied from Microsoft Excel
% Workbook: /Users/haiiro/Dropbox/Nearsightedness:Tables.xlsx
% Worksheet: Sheet1
% Range: $Y$17:$AK$18

\begin{table}[h!]
	\centering
	\caption{\label{table:Si-summary}
		Si atomic diameter, energy of formation, isolated atomic energy, and changes in central Bader atom energy resulting from the addition of cluster coordination spheres ($\bm{\Delta E^x_n}$) as described in the text.
		Distances are reported in \AA\ and energies in eV.
		$\bm{\Delta E_{10}}$ was not determined.
	}
	\begin{tabular}{|c|c|c|c|c|c|c|c|c|c|c|c|c|}
		\hline
		\textbf{Si diameter (\AA)}
		& $\bm{E_f}$
		& $\bm{E_0}$
		& $\bm{\Delta E_1}$
		& $\bm{\Delta E_2}$
		& $\bm{\Delta E_3}$
		& $\bm{\Delta E_4}$
		& $\bm{\Delta E_5}$
		& $\bm{\Delta E_6}$
		& $\bm{\Delta E_7}$
		& $\bm{\Delta E_8}$
		& $\bm{\Delta E_9}$
		& $\bm{\Delta E_{11}}$
		\\\hline
		2.352
		& -5.42
		& -7865.72
		& -3.05
		& -9.01
		& -5.66
		& -5.55
		& -6.09
		& -6.71
		& -6.67
		& -6.12
		& -5.91
		& -5.79
		\\
		\hline
	\end{tabular}
\end{table}

Silicon possesses the diamond cubic crystallographic structure. 
Its near neighbor shell structure is summarized in \cref{table:DC}, with the central crystalline region represented by a five atom cluster (\ch{Si5}) composed of a single atom and its four tetrahedrally coordinated nearest neighbors.

% DC 7th, BCC 5th (59), FCC 5th, HCP 5th
Relative to the energy of an isolated Si atom, $E_0$, the calculated per atom Bader energy of \ch{Si5} as its boundary region was increased to include a second, third, fourth and so on up to eleven coordination shells are reported in \cref{table:Si-summary} along with the nearest neighbor distance of Si and its calculated formation energy, $E_{\!f}$.

\begin{figure*}[h]
\centering
%\vspace{-25 pt}
\includegraphics [width=0.9\linewidth]{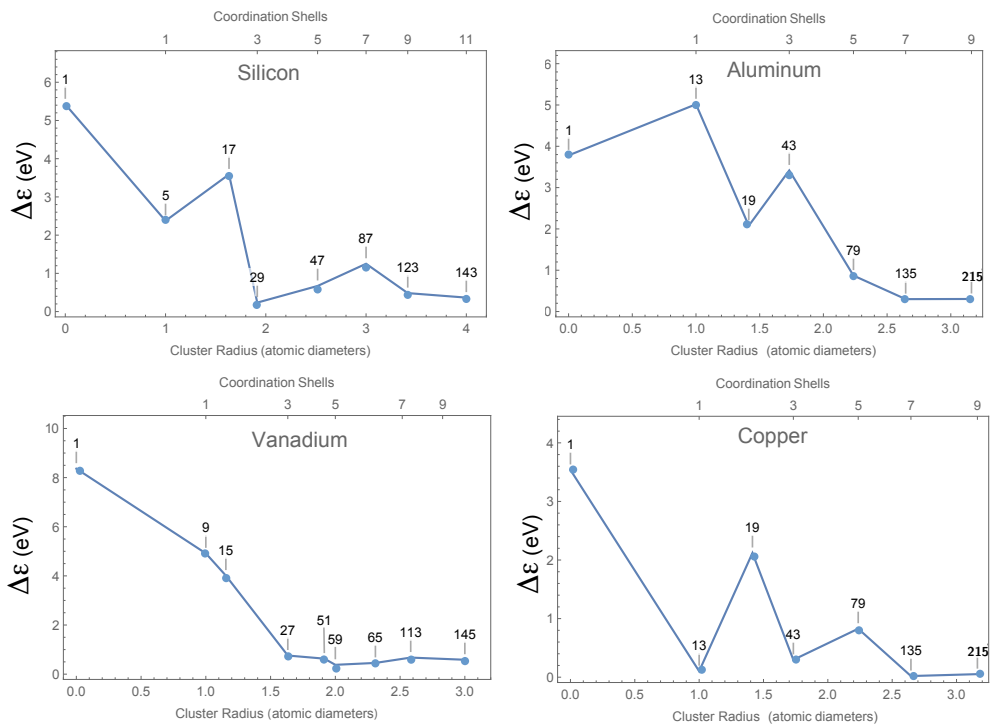}
%\begin{singlespace}
%\vspace{-10 pt}
\caption{\label{fig:groupI_energies}
	The nearsightedness function ($\Delta \varepsilon^x_n$) for the elements Al, Si, V and Cu as a function of $n$.
	The callouts in the graphs give the number of atoms in the representative cluster.
	For example, an Al cluster representing a central atom and its first 7 coordination spheres will contain 135 atoms.
}
%\end{singlespace}
\end{figure*}

These results are summarized graphically in the upper left of \cref{fig:groupI_energies} where $\Delta \varepsilon^{\text{Si}}$ is depicted as a function of $n$.
This figure represents the sensitivity of the central cluster to the retreating perturbation, that is, the distance over which the central cluster can clearly see the free surface. When the energy goes to zero, from the central cluster's point of view, the surface has vanished. Plainly, when the free surface is infinitely distant from the central cluster, the cluster's per atom energy will be identical to that of crystalline Si. Hence beyond some point, to computational accuracy, the decay of $\Delta \varepsilon$ will approach zero asymptotically. In fact, because for both ordered and disordered gapped materials---the systems modeled here are all gapped, in that the central cluster energy converges to within computational accuracy before the energy difference between the LUMO and HOMO is on the order of $k$T---the change in the density due to a perturbation  at $R$ decays exponentially with $R$,\cite {NEM_PNAS} it is arguable that energy decay should not only be asymptotic but exponential.

Regardless, inspection of \cref{fig:groupI_energies} reveals that the onset of the asymptotic/exponential decay begins with coordination shell seven, where the per atom energy difference between crystalline silicon and the central cluster is on the order of an eV, decreasing to 0.36 eV at eleven coordination shells---three atomic diameters beyond the central cluster.

Naturally, this variation of $\Delta \varepsilon$ is due to the changing boundary width and its associated influence on the central cluster charge density.\cite{molecules_in_metals}
And just as the energy of a central cluster with an infinite boundary will be equivalent to that of a crystal, so too will the charge density of a cluster with an infinitely wide boundary be identical to the crystalline density. 

For all elemental crystals, equivalence between the central cluster and crystalline charge densities is required when the Bader atom surfaces of the central cluster are coincident with the crystalline Voronoi polyhedra (cells) about each atom.
Quite generally, the difference between the surface of a cluster's Bader atoms and a crystal's Voronoi cells provides a measure of their charge density differences, which vanish when the two surfaces coincide.\cite{holographic} 

\begin{figure}%{r}{0.25\textwidth}
%	\vspace{-30 pt}
	\begin{center}
		\includegraphics[width=0.5\linewidth]{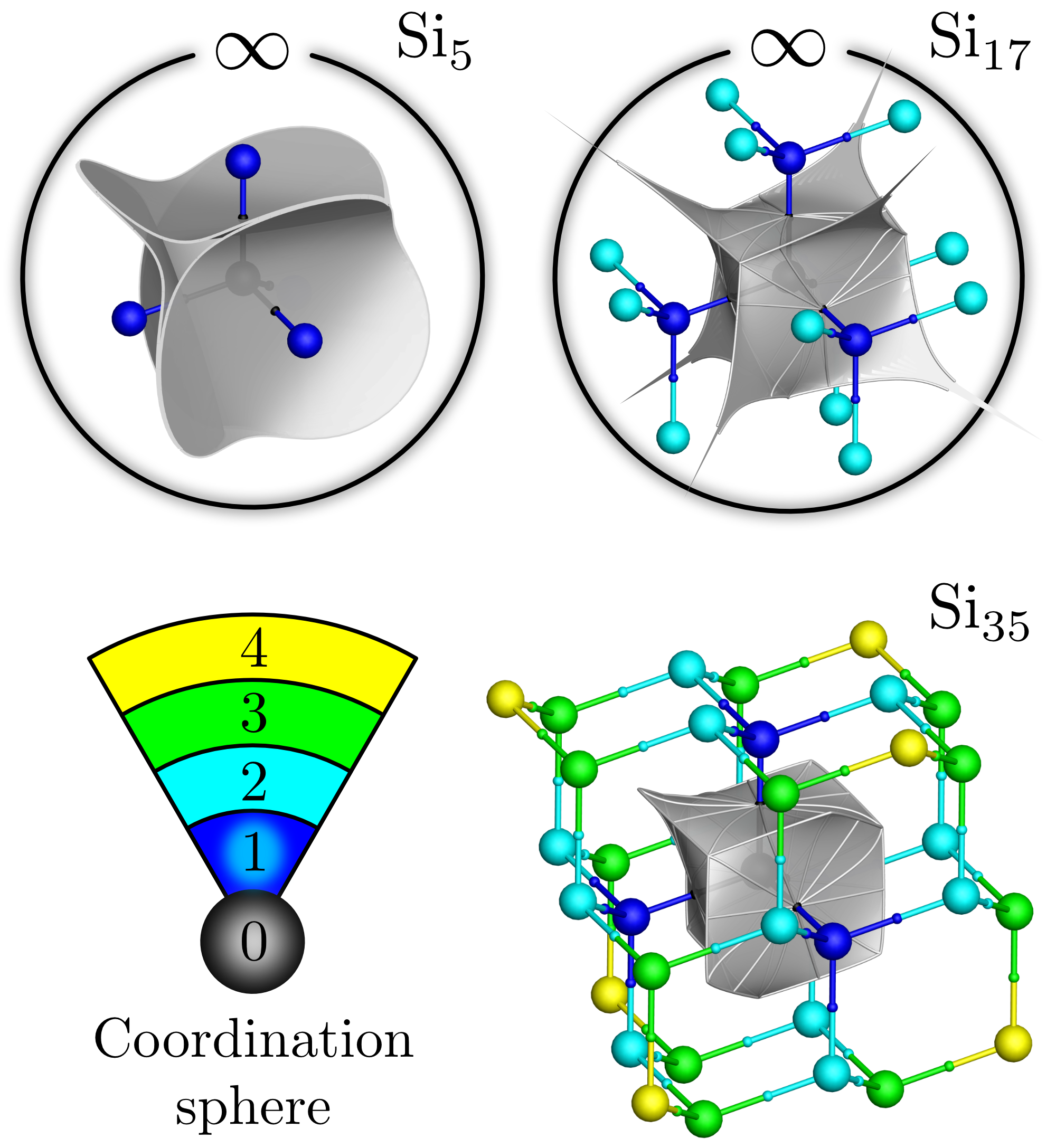}
	\end{center}
%	\begin{singlespace}
%		\vspace{-30 pt}
		\caption{\label{fig:Si_bader_atoms} Depiction of the central Bader atom surfaces in 5, 17, and 35 atom clusters (1, 2, and 4 coordination spheres respectively).
		Nuclear positions are indicated by spheres colored according to coordination sphere.
		In \ch{Si5} (top-left) the central Bader atom remains clearly open in four regions, one of which is facing to the left of center, while elsewhere the Bader atom surfaces are essentially converged.
		With two coordination spheres (top-right; \ch{Si17}) those same open regions have closed coincident with topological cage points, and only slivers of very-nearly converged surfaces prevent the Bader atom from being closed.
		At four coordination spheres (bottom-right; \ch{Si35}) the Bader atom has closed completely.
		The open surfaces in \ch{Si5} were truncated according to the $0.001\text{e}^-$ charge density isosurface.}
%	\end{singlespace}
%	\vspace{-10 pt}
\end{figure}

A Bader atom's surface must contain local charge density minima.
(In the chemical literature \cite{AIM} these minima are called cage critical points to indicate that there is one such point interior to cages of bound atoms.) Plainly, an atom's local charge density minima may be a finite or an infinite distance from the atomic nucleus.
If all the local minima are a finite distance from the nucleus, the surface of the Bader atom is topologically connected and the atom is said to be closed.
On the other hand, if even one local minimum is located at infinity, the Bader atom surface is disconnected and the atom is said to be open.
Importantly, for any open Bader atom there is a path lying entirely within the atom that runs from the location of the nucleus to a point at infinity.
Quite simply, an open atom is characterized by a channel of charge density connecting the nucleus to the neighborhood of at least one infinitely distant point.
In contrast, the surfaces of crystalline Voronoi polyhedra are necessarily connected.
For example, the Voronoi cell of the diamond cubic structure is in a class of truncated tetrahedra. 
% triakis truncated tetrahedra having sixteen faces, four hexagonal and 12 triangular thirty edges and sixteen vertices. 

As a means of clarifying this issue, consider the evolution of a central region's Bader atoms as a cluster grows.\cite{molecules_in_metals}
In the case of Si, this process is represented in \cref{fig:Si_bader_atoms}.
The top-left frame depicts the Bader atom surfaces of \ch{Si5}, or equivalently the interatomic boundaries between the central atom and its first coordination sphere.
This set of surfaces is constructed from four asymptotic---hence disconnected---surfaces.
As a result, all  the Bader atoms of the \ch{Si5} cluster are open.
In other words, around every point there is a direction in which the charge density is decreasing and thus local minima are infinitely distant from the central atom. 

The boundary of the central Bader atom evolves with the addition of the twelve atom second coordination sphere to yield the \ch{Si17} cluster pictured in the top-right frame of \cref{fig:Si_bader_atoms}.
While the asymptotes separating the surfaces of this atom become steeper, the central Bader atom still extends to infinity.
In other words, from some points near the central atom there is a path of decreasing charge density that leads to infinity.
This path will be located within the ``spikes'' evident in the top-right frame of \cref{fig:Si_bader_atoms}. 

As depicted in the bottom-right frame of \cref{fig:Si_bader_atoms}, it is with the addition of the fourth coordination sphere to make a \ch{Si35} cluster that the boundary of the central Bader atom is topologically connected and the atom is closed.
It is at this point that the cages of bound atoms sharing the central atom as a common vertex are completed, which mandates a single local minimum at the center of each of these cages.
In a sense, it is with the completion of these cages, and the resultant closing of its Bader atom, that the central atom becomes isolated from the surroundings through an intervening shell of charge density. 

While the central atom is closed with the fourth coordination shell, the central cluster closes at the seventh coordination shell, {\it i.e.}~\ch{Si87}.
It is here that the cages having a first coordination sphere atom as a vertex are completed.
In fact, at this point the Bader atoms of the central cluster possess the topology of the Voronoi polyhedra of crystalline silicon.
And, though \cref{fig:groupI_energies} shows only a few points beyond the seventh coordination shell, $\Delta \varepsilon$ decreases monotonically (arguably exponential decay) through these points.	

Returning to the remaining crystals of the first set:~Al, V, and Cu.
The central cluster of the FCC metals (Al and Cu) is a cuboctahderon consisting of a central atom and its twelve nearest neighbors.
The central cluster of BCC V is a nine atom cube with an atom at the cube center and its eight nearest neighbors located at the cube vertices.
Pictures of these clusters  along with analogues of \cref{table:DC} giving more information regarding the shell structure of FCC and BCC crystals along with the element specific cluster energy analogues of \cref{table:Si-summary} are provided in the SI.

This information is also summarized graphically in \cref{fig:groupI_energies} where $\Delta \varepsilon_n$ for the central cluster of each element is plotted as a function of $n$.
In all cases, within ten coordination spheres $\Delta \varepsilon$ converged to within a fraction of an eV of the computed crystalline formation energy ($\Delta \varepsilon^{\text{Al}}_{9},\; \Delta \varepsilon^{\text{V}}_{10}, \; \Delta \varepsilon^{\text{Cu}}_{9} = $ 0.30, 0.58, 0.05 eV respectively).
More noteworthy however, though the onset of exponential decay is element dependent---2, 3, and 5 coordination spheres for V, Al, and Cu respectively---beyond the point where the Bader atoms of the central cluster close (BCC clusters of 59 atoms and FCC clusters of 79 atoms) $\Delta \varepsilon$ is monotonically decreasing or level.

\subsection{4\textit{d} Metals}

\begin{figure*}[h]
\centering
%\vspace{-25 pt}
\includegraphics [width=0.95\linewidth]{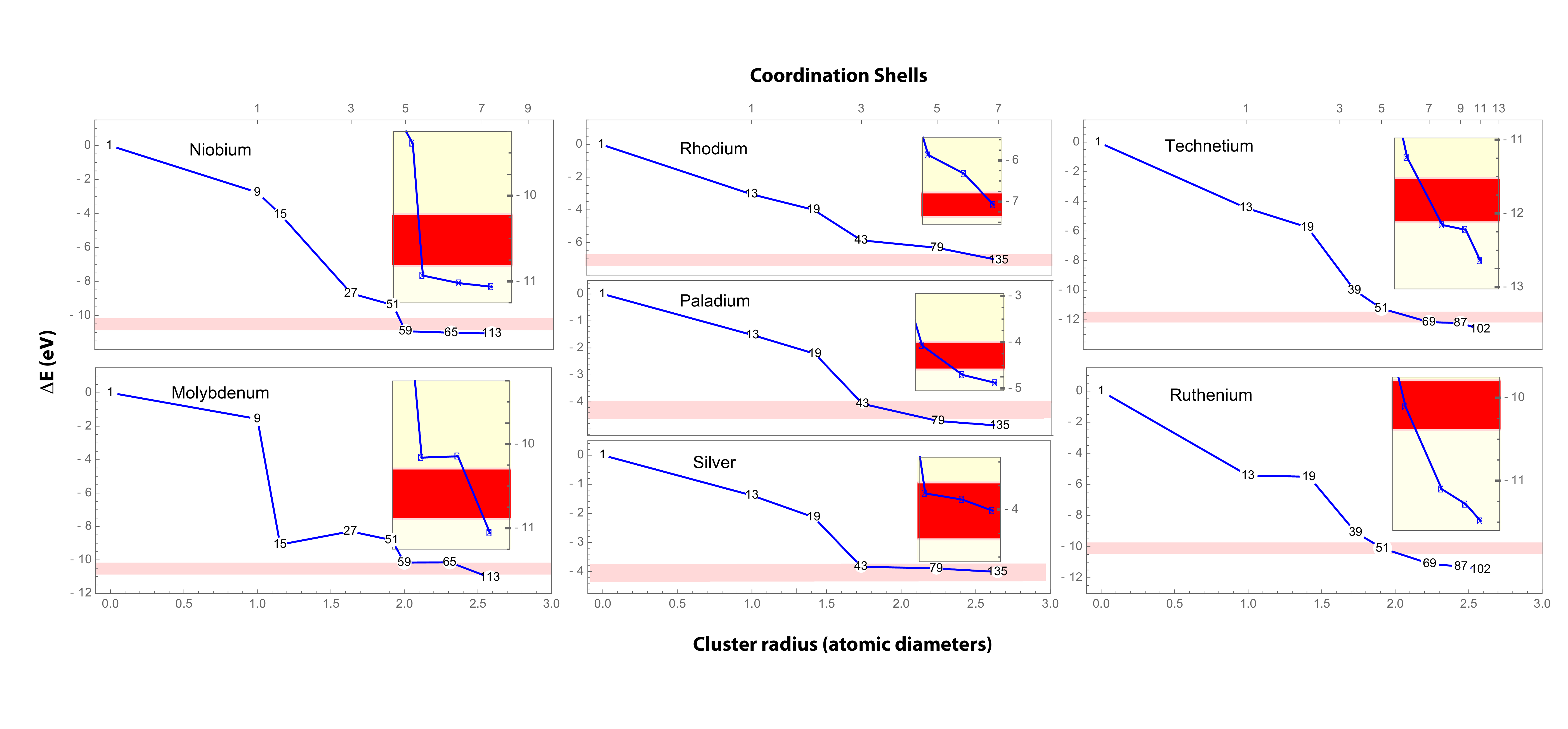}
%\begin{singlespace}
%\vspace{-10 pt}
\caption{\label{fig:group2_energies}
	The change in central Bader atom energy resulting from additional cluster coordination shells ($\Delta E$) for the BCC (left column), FCC (center column) and HCP (right column) 4$d$ transition metals.
	The red stripes indicate the crystalline formation energy assuming a $\pm0.3$ eV basis set error.
	Insets provide a higher fidelity depiction of $\Delta E$ for the three largest clusters of the series.}

%\end{singlespace}
\end{figure*}

The heavier 4$d$ transition metals required a larger basis set for the accurate determination of Bader atom energies. Unfortunately, the use of this larger basis set introduced an estimated 0.6 eV imperscision (see SI) to the calculated values of the crystalline formation energy.  Nonetheless, 
$\Delta E$ as a function of the number of coordination shells for the BCC metals Nb and Mo; the FCC metals Rh, Pd and Ag; and the HCP metals Tc and Rh are shown in \cref{fig:group2_energies}.
The information contained in these figures is provided in tabular form in the SI.

%\break

Like the first set of crystals, all metals appear to be converging asymptotically on a value near their estimated formation energies for clusters with radii of 2.5 atomic diameters.
More significant however, across a given structure type--BCC, FCC, HCP---the form of the decay of $\Delta E$ is similar.
This is particularly noticeable for the BCC metals Nb and Mo.

Recall the central nine atom cluster of a BCC metal closes at 59 atoms or equivalently five coordination spheres.
\Cref{fig:group2_energies} reveals a rapid decrease of more than an eV in the central cluster energy on addition of the fifth coordination sphere---transforming a 51 to a 59 atom cluster;
and for both Nb and Mo bringing the central cluster energy to within computational accuracy of the estimated formation energy.
While in V the effect is not as dramatic, a similar central cluster stabilization was observed (\cref{fig:groupI_energies}) with the addition of the fifth coordination sphere.

The decay of $\Delta \varepsilon$ for the FCC metals as shown in the center column of \cref{fig:group2_energies} also exhibit similar forms.
The most pronounced energy change accompanies the  closing of the central Bader atom with the addition of the second coordination sphere, which also marks the onset of exponential decay. The closing of the central cluster at five coordination spheres and 79 atoms is well inside the ``asymptotic'' region of the decay and  is not as sharply defined as in the 4$d$ BCC metals. 

Unlike the BCC metals, characterized by a single type of local minimum, the FCC metals have two; the first at the center of the FCC octahedral hole, and the second at the center of the FCC tetrahedral hole.
Some of the channels linked to these minima are plugged earlier than are others. The tetrahedral holes of the central atom form with the addition of the first coordination sphere and the central atom octahedral holes with the second coordination sphere.  
The addition of the third coordination sphere at a cluster size of 43 atoms, which correlates with a substantial central cluster stabilization, begins closing the atoms in the first coordination sphere.
Nonetheless, the central cluster fully closes only with the addition of the fifth coordination shell.
And it is at this size that the energy of the central cluster falls to within 0.15 eV of the estimated formation energy for all the 4$d$ FCC transition metals.

 There are 4 symmetry unique local minima types in HCP clusters. 
Like the FCC structure there are both tetrahedral and octahedral minima. And like the FCC metals, the onset of exponential decay   begins with the addition of the second coordination sphere.
However, with the central cluster possessing D$_{3h}$ symmetry, the octahedral and tetrahedral holes  are split into symmetry unique pairs depending on their displacement perpendicular or parallel to the 3-fold axis.
The central thirteen atom cluster undergoes significant closure with the third coordination shell and completely closes with the fifth coordination shell---a cluster of 51 atoms. 

Thus, across  all structure types modeled,  the energy of the central cluster ``improves'' in lock step with the progressive closing  of the central cluster. The underlying factors driving  this observation are   rooted in the virial theorem.\cite{Slater_virial, Rudenberg_virial}

When extended to Bader atoms \cite{AIM} the virial theorem establishes that for an atom $\Omega$ at mechanical equilibrium, its average kinetic energy $ \langle T_\Omega \rangle $, and its average potential energy
$ \langle V_\Omega \rangle $ are related by,
$$ \langle V_\Omega \rangle = -2\langle T_\Omega \rangle $$
and hence its total energy $E_\Omega$ is simply,
$$ E_\Omega = -\langle T_\Omega \rangle.$$
Ergo, as an atom lowers its total energy through interactions with its surroundings, its kinetic energy must simultaneously increase.
However, the virial theorem applies only to average energies, therefore the regions of increased kinetic energy need not coincide with the regions of decreased total energy. 

Bader  argued that the ``quantum mechanically local kinetic energy'' at a point $\bm{r}_0$, is related to the magnitude of $\nabla^2 \!\rho(\bm{r}) |_{\bm{r}_0}$.\cite{Bader_kinetic_energy}
Hence, deep, steeply curved minima contribute positively to the kinetic energy and allow electron rearrangement in other regions that will lower $ E_\Omega$.
Also we note that as charge density minima lie along interatomic nodes of the one-electron wave functions, they may coincide with anti-bonding interactions where kinetic energy is high.\cite{wilson2019}
%Plainly, charge density minima, and the network of linked atoms that produce these minima should be considered as an inseparable unit, {\sl i.e.}~as a neighborhood. 

\subsection{Nearsightedness and Embrittlement} 

We conclude that  ``good''  models  of crystalline environments will at a minimum envelop a neighborhood that extends to local charge density minima and thereby recover the topology of $\rho(\bm{r})$.
It is worth considering the fundamentals driving  this observation.  

In the cases we have considered, charge density minima and the cages they represent result from  $p$- and  $d$-orbital alignments  in which the crystalline environment couples  $\sigma$-, $\pi$- and $\delta$- interactions so as to minimize {\bf global energy}.  
Models that exclude charge density minima effectively decouple the orbital interactions to allow {\bf local energy} minimization and hence cannot serve as good crystalline representations.  
However, while  decoupling in crystals is an artificial effect stemming from an inadequate model, it is a real effect around defects, possibly altering the nearsightedness function and the properties mediated by nearsightedness.  
One such property is impurity induced  embrittlement.     

Impurity induced  embrittlement  often begins with the segregation of dilute impurities to grain boundaries, which, through a not entirely understood mechanism, yields them susceptible to brittle failure.   
Because this  phenomena is of great economic consequence, and is often associated with catastrophic failure leading to the loss  of life,\cite{failure1,failure2} it has been extensively studied.  
Particularly significant are studies seeking to correlate embrittlement with changes to electronic structure caused by segregation. These studies have  resulted in three proposed mechanisms.  
The first, posits that impurity atoms weaken the boundary by withdrawing  electron density from the cohesive metal-metal bonds.\cite{messmerbriant}  
The second is a thermodynamic model implicating the   difference between grain boundary and  surface energy as the parameter controlling embrittlement potency.\cite{geng,wu2}  
According to this model, the more an element lowers surface energy relative to grain boundary energy, the greater its embrittling potential.
The third model attributes embrittlement to changes in the directionality of the grain boundary bonds.  
In one instance, it has been argued that embrittling elements make intergranular bonds more directional,\cite{haydock1981mobility} and in another  less directional.\cite{eberhart} 
However, none of these consider possible synergistic effects between the  sites of fracture initiation and  embrittling atoms. 

All fractures begins in the vicinity of a stress concentrator, conventionally thought of as an atomically sharp crack.\cite{griffith1, griffith2}   
The importance of such cracks cannot be discounted.  Ductile substances may fail in an apparently brittle fashion through the introduction of a sufficiently sharp crack.   On the other hand, normally brittle materials can be rendered deformable by eliminating surface  cracks.\cite{joffe}
Traditionally such cracks are envisioned as having an easily identifiable  tip, with bonds on  the crack side broken and  unable to carry  load.  An applied stress normal to the crack will of necessity concentrate  in the unbroken bonds on the other side of the  tip, causing them to preferentially elongate, ultimately break and thereby growing the crack.  Elasticity theory attributes the stress concentration from a classical crack to its length  and the radius of its tip, neither of which has meaning in a real material where instead  energy localization derives from the crack tip charge density.  
%We speculate that embrittling elements make  crack tips more nearsighted and hence  more effective stress concentrators and explore the bismuth doped copper system in support of this speculation.

Owing to its comparative simplicity, bismuth doped copper serves as one prototypical system for the investigation of intergranular embrittlement.\cite{cubi1, cubi2, cubi3, cubi4, cubi5}   
And consequently  a wealth of data  has been generated on this system through both theoretical and experimental investigations.\cite{cubi3, cubi5, cubiexp1, cubiexp2, cubiexp3, grainboundary, cubiexp4, losch}  
Particularly important to our efforts is the study by Duscher {\it et al.},\cite{grainboundary}  which included atomic resolution Z-contrast images of the grain-boundary region of a symmetric 36.8$^\circ$ <001> tilt boundary known to be embrittled through Bi segregation.  

\begin{figure}[h]
	\begin{center}
		\includegraphics [width=0.6\linewidth]{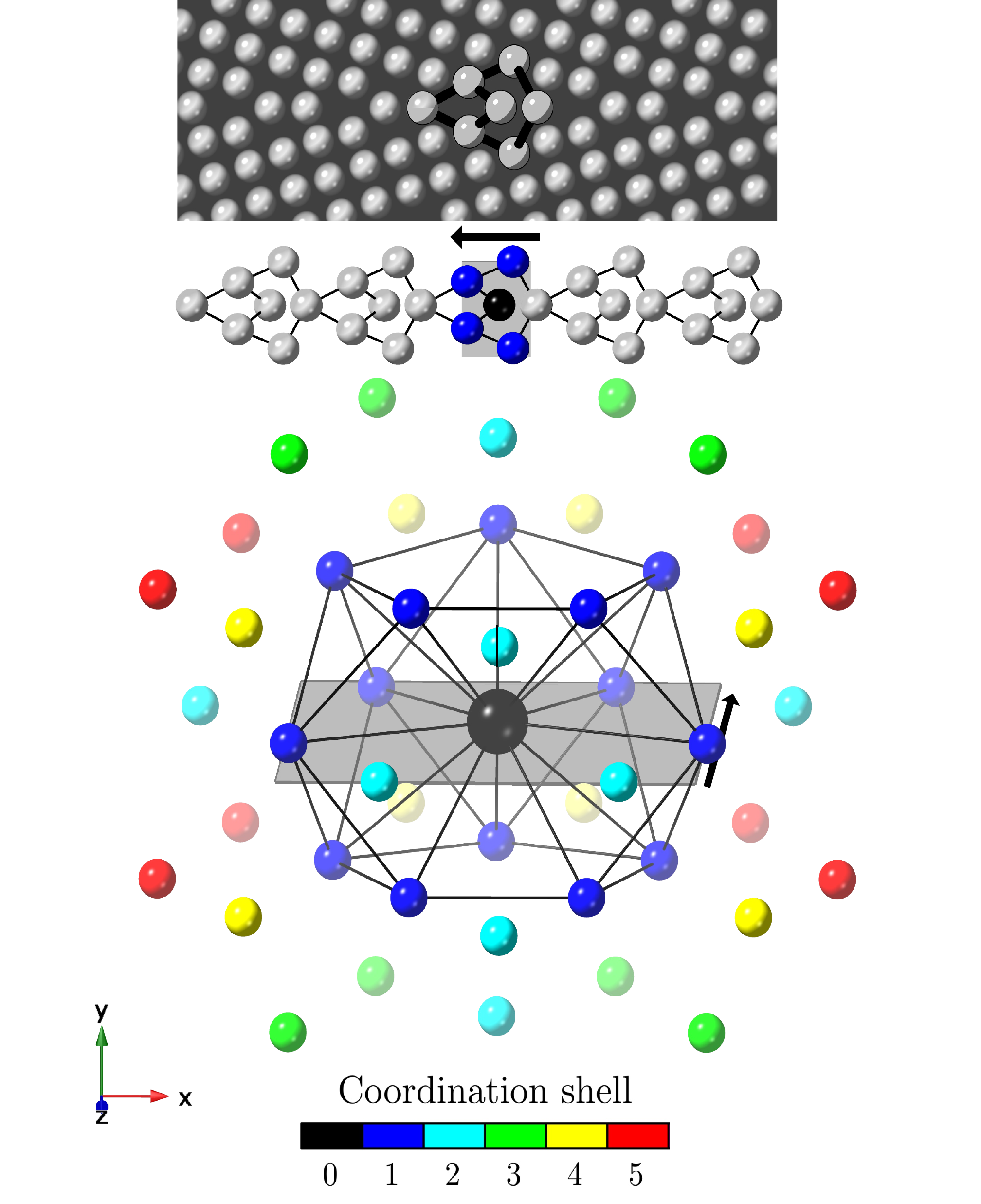}
	\end{center}

		\caption{\label{fig:Bi_Cu} Depiction of the Bi-Cu boundary cluster. A simulated image of the grain boundary region from reference \citenum{grainboundary} with the structural unit is indicated (top). The 2-dimensional repeating structural unit (middle), and the corresponding 3-dimensional cluster with the Bi atom in the center (bottom) are shown. The atoms are color coded based on coordination sphere; and the central cluster is represented by the first coordination sphere. The XZ plane, indicated by the gray plane in the bottom frame coincides with the grain boundary plane and corresponds to the grey shaded rectangle in the middle frame, where the same atom coloring is used to show the central cluster atoms in the boundary. }

\end{figure}

The structure of this boundary is shown in \cref{fig:Bi_Cu} and is characterized by a repeating kite structure.  Bi was found to substitute for the Cu atom at the center of this kite  where it sits in a roughly pentagonal coordination shell of 14 atoms, also shown in \cref{fig:Bi_Cu}.  We take this 14 atom shell containing a central Bi or Cu atom as the central cluster of our nearsightedness investigation.  

\begin{figure}[h]
	\begin{center}
		\includegraphics [width=0.7\linewidth]{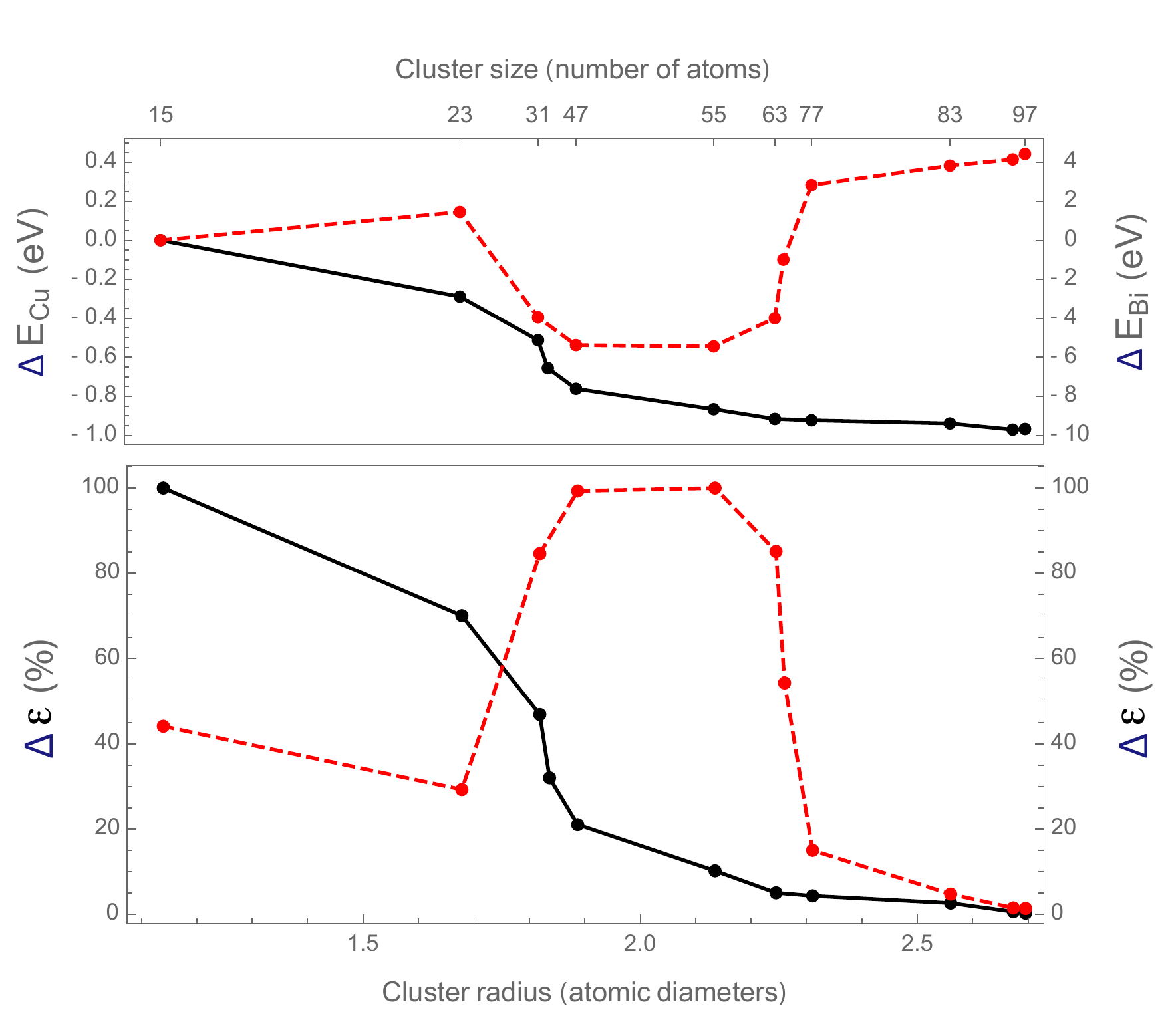}
	\end{center}
		\caption{\label{fig:Bi_Cu_energies}
		Change in central cluster Bader energy resulting from additional grain boundary cluster coordination shells.
		\textbf{Top:} Per atom $\Delta E$ for Cu (dashed red) and Bi (solid black) centered grain boundary clusters.  
		\textbf{Bottom:}    $\Delta \varepsilon$ normalized by the respective maximum magnitudes of the perturbation energies.}
\end{figure}

As in the crystalline studies, we determined the change to the energy of this cluster  through the addition of successive coordination spheres. The calculated energies are given in the SI and the results are shown graphically in  \cref{fig:Bi_Cu_energies}, where the top frame gives the raw values for the two---Cu and Bi containing---clusters.  The bottom frame of the figure gives the ``normalized'' values of $\Delta \varepsilon_n$, where the maximum value of $\Delta \varepsilon_n$ is used as a normalizing factor.  This bottom frame allows one to compare  the response of pure Cu and Bi segregated systems  to perturbation from  successive bicrystalline shells, or conversely,  how  a perturbation to the  central cluster containing either Cu or Bi is distributed to the surrounding bicrystal. 

Unlike the crystalline systems, the central grain boundary cluster of pure Cu becomes less stable  due to the perturbing influences of the surrounding environment. This behavior results from competition between interactions that lower either global or local energy.    As mentioned, interactions that promote global stabilization of the FCC structure are a consequence  of aligning Cu $d$-orbitals so as to maximize the combined contributions from $\sigma$-, $\pi$- and $\delta$-overlap.  Global energy minimization is highly constrained and hence the interactions responsible for this stabilization are termed  directional. Local stabilization results from charge transfer between an atom and its immediate environment  to bring $s$-orbitals into energy alignment and promote non-directional  coulomb attraction.   

For the isolated  central grain boundary cluster, non-directional interactions predominate.  The  Cu atom at the center of the non-crystallographic 14 atom shell develops a Bader charge of 0.24 electrons  that is necessarily transferred to the 14 atom shell---Hirshfield charge density analysis shows the same charge transfer.
As successive shells are included in the model, the charge transfer stays essentially constant, dropping to 0.21 electrons through the 63 atom cluster, but decreasing more rapidly from that point on, reaching a value of 0.12 electrons at the 97 atom cluster.  

Inspection of \cref{fig:Bi_Cu_energies} reveals a corresponding steep change to the perturbation energy at the 63 atom cluster and a distance of roughly 2.2 Cu-Cu near neighbor separations.  Remarkably, this is very near the  distance at which the central cluster of  crystalline Cu  (see \cref{fig:groupI_energies}) fully closes with an accompanying  onset of exponential energy decay.  Hence beyond 2.2 Cu diameters, the perturbation to the central grain boundary cluster due to the surrounding bicrystalline environment exceeds the stability gained from the non-directional coulomb interactions.   The charge density  of the central cluster responds by aligning  its $d$-orbitals with those of the grains on either side of the boundary, which reduces the energy of distant atoms.  However,  because the $d$-orbitals of the central cluster cannot be internally aligned,  due to the grain boundary  misorientation, the central cluster  energy increases.  Overall though, there is a net stabilizing effect to the system, which decays exponentially beyond 2.2 Cu diameters and becomes insignificant relative to the grain boundary energy beyond 2.5 Cu diameters.  That is, movement of the atoms at the core of a pure Cu grain boundary are felt across a sphere of approximately 2.5 Cu diameters. 

The Bi containing central cluster behaves similarly to the pure Cu cluster out to roughly 2.1 Cu diameters.  But unlike  pure Cu, the energy of the central cluster does not change substantively  with increasing coordination spheres of bicrystalline environment.     This response is due to the valence $p$-orbitals on the Bi atom, which combine with Cu orbitals to form localized states that inhibit Cu atom $d$-orbitals on the central cluster from aligning with those on the adjacent grains.  Effectively, the Bi containing boundary is more nearsighted than the pure Cu boundary and cannot see the bicrystalline environment beyond 2.1 (arguably 1.9) Cu diameters.

Consider now  a load applied normal to an atomic scale crack that at one end intersects a Cu grain  boundary.  This crack will act as a stress concentrator.   On the one hand, in the absence of segregated  impurities, perturbations to the positions of the crack tip atoms resulting from the applied load will be distributed to a significant extent over a spherical volume with a radius of about 2.5 Cu diameters.  On the other hand, in the presence of segregated Bi atoms, the same load induced perturbations will be distributed  across a sphere with a radius of about 2.1 Cu diameters, a neighborhood with a volume about  60\% as large as that of the impurity free boundary.  For all practical purposes,  Bi atoms  sharpen Cu crack tips.

In the Cu-Bi system we hypothesize that embrittlement results from  impurity induced nearsightedness and hence an increase to the crack tip stress intensity factor.  In effect, nearsightedness is the quantum mechanical counterpart to  continuum mechanics' crack tip radius.  As such, increased crack tip nearsightedness, whether induced by impurities or the crack tip atomic structure itself, is also a necessary and perhaps sufficient condition for brittle fracture.  Though here the mechanism responsible for enhanced nearsightedness is the result of greater non-directional bonding, resulting in local elastic constant softening---consistent with the findings of references  \citenum{eberhart} and \citenum{grainboundary}---we see no reason to believe that this is the sole mechanism  responsible for increased nearsightedness.  More thorough investigations of nearsightedness and its underlying atomic origins may shed light  not only on environmentally induced embrittling phenomena, where the composition of embrittling elements is often sharply peaked around  stress concentrators, but also on intrinsic brittleness and the host of phenomena involving mechanical and chemical energy focussing and localization such as: explosive hot spot formation,\cite{hot_spot}  triboluminescence, \cite{triboluminescence} sonoluminescence, \cite{sonoluminescence} sonochemistry, \cite{sonochemistry} and enzyme electrostatic preorganization. \cite{electro_preorg}

\section{Summary}

%In conclusion: 
We have defined a nearsightedness function giving the local perturbation energy due to an increasingly distant free surface. This function allows one to identify the neighborhoods that may serve as models of crystalline and defected materials. The boundaries of these neighborhoods are determined by the topological properties of the charge density and were found to be a few atomic diameters in radius.  We argued that these neighborhoods serve as adequate models for the study of many properties and particularly mechanical properties of metals. We supported this argument by demonstrating that Bi atoms segregated to a Cu grain boundary  increase nearsightedness and suggested that  enhanced nearsightedness manifests  most prominently through  increases to  the crack tip stress intensity factors.  We further proposed that the nearsightedness function may prove useful in investigations of diverse phenomena involving energy focussing and localization.

\section*{Acknowledgments}

The authors would like to acknowledge the contribution of Dr.\ Garritt Tucker and Jacob Tavenner from the Computational Materials Science and Design group at the Colorado School of Mines for providing  grain boundary structures.
We also acknowledge valuable discussions with Dr.\ Travis Jones from the Fritz Haber Institute of the Max Planck Society.
Support of this work under ONR Grant No.~N00014-10-1-0838 is gratefully acknowledged. 

\singlespacing
\section*{References}
\bibliography{Refs}



\section*{Supplementary material}

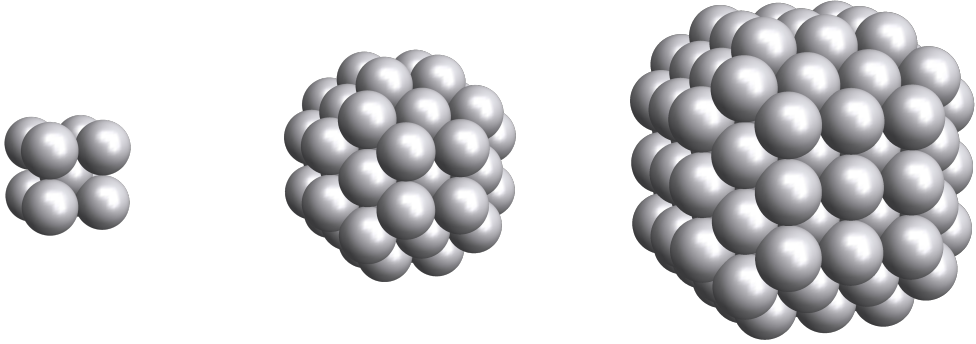

Figure 2: The BCC central 9-atom cluster (left), critical cluster consisting of 5 coordination shells (middle), and the largest cluster consisting of 10 coordination shells (right).

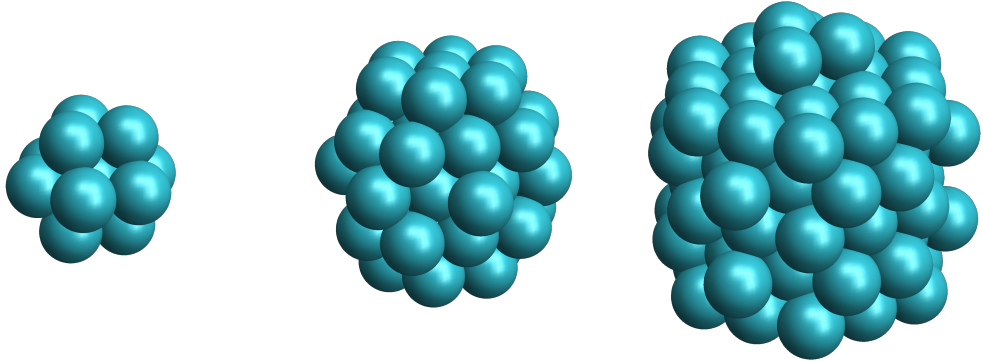

Figure 3: The HCP central 13-atom cluster (left), critical cluster consisting of 5 coordination shells (middle), and the largest cluster consisting of 11 coordination shells (right).

Table 4: Diamond cubic (DC) shell structure. Table layout is identical to that of Table 1.

| Coordination shell $n$ | 0 | 1 | 2 | 3 | 4 | 5 | 6 | 7 | 8 | 9 | 10 | 11 |
|---|---|---|---|---|---|---|---|---|---|---|---|---|
| Number of $n^{\text{th}}$ neighbors | 0 | 4 | 12 | 12 | 6 | 12 | 24 | 16 | 12 | 24 | 12 | 8 |
| Total atoms in cluster | 1 | 5 | 17 | 29 | 35 | 47 | 71 | 87 | 99 | 123 | 135 | 143 |
| Cluster radius | 0 | 1 | $2\sqrt{\frac{2}{3}}$ | $\sqrt{\frac{11}{3}}$ | $4\sqrt{\frac{1}{3}}$ | $\sqrt{\frac{19}{3}}$ | $2\sqrt{2}$ | 3 | $4\sqrt{\frac{2}{3}}$ | $\sqrt{\frac{35}{3}}$ | $\sqrt{\frac{43}{3}}$ | 4 |

## C  Element data

Data from which the graphs shown in the main text were constructed are provided in Tables 5-8.

Table 5: Perturbation energies $\Delta E$ of the (single) DC cluster investigated, reported in eV.

| Element | Diameter (Å) | $E_f$ | $E_0$ | $\Delta E_1$ | $\Delta E_2$ | $\Delta E_3$ | $\Delta E_4$ | $\Delta E_5$ | $\Delta E_6$ | $\Delta E_7$ | $\Delta E_8$ | $\Delta E_9$ | $\Delta E_{11}$ |
|---|---|---|---|---|---|---|---|---|---|---|---|---|---|
| Si | 2.352 | -5.42 | -7865.72 | -3.05 | -9.01 | -5.66 | -5.55 | -6.09 | -6.71 | -6.67 | -6.12 | -5.91 | -5.79 |

c

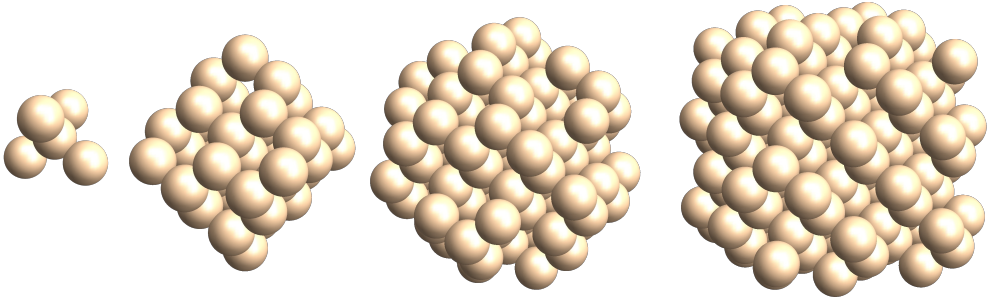

Figure 4: (from the left) The DC central 5-atom cluster, the 4-coordination cluster closing the central atom, the 7-coordination cluster closing the central cluster, and the largest cluster consisting of 11 coordination shells.

Table 6: Perturbation energies $\Delta E$ of FCC clusters, reported in eV.

| Element | Diameter (Å) | $E_f$ | $E_0$ | $\Delta E_1$ | $\Delta E_2$ | $\Delta E_3$ | $\Delta E_4$ | $\Delta E_5$ | $\Delta E_6$ | $\Delta E_7$ | $\Delta E_9$ |
|---|---|---|---|---|---|---|---|---|---|---|---|
| Al | 2.864 | -3.78 | -6587.00 | -8.80 | -5.86 | -7.20 | - | -4.65 | - | -3.48 | -4.08 |
| Cu | 2.553 | -3.54 | -44611.75 | -3.64 | -5.66 | -3.83 | - | -4.37 | - | -3.56 | -3.56 |
| Rh | 2.758 | -6.78 | -127504.31 | -3.05 | -4.00 | -5.87 | - | -6.32 | - | -7.08 | - |
| Pd | 2.751 | -3.99 | -134362.32 | -1.52 | -2.22 | -4.07 | - | -4.71 | - | -4.88 | - |
| Ag | 2.885 | -3.71 | -141429.58 | -1.38 | -2.13 | -3.83 | - | -3.90 | - | -4.01 | - |

Table 7: Perturbation energies $\Delta E$ of BCC clusters, reported in eV.

| Element | Diameter (Å) | $E_f$ | $E_0$ | $\Delta E_1$ | $\Delta E_2$ | $\Delta E_3$ | $\Delta E_4$ | $\Delta E_5$ | $\Delta E_6$ | $\Delta E_7$ | $\Delta E_8$ | $\Delta E_{10}$ |
|---|---|---|---|---|---|---|---|---|---|---|---|---|
| V | 2.624 | -8.36 | -25665.42 | -3.45 | -4.35 | -7.60 | -8.99 | -8.74 | -8.82 | -9.33 | -9.04 | -8.95 |
| Nb | 2.857 | -10.22 | -102139.80 | -2.77 | -4.06 | -8.71 | -9.39 | -10.93 | -11.02 | -11.00 | -11.06 | - |
| Mo | 2.725 | -10.29 | -108177.90 | -1.57 | -9.08 | -8.29 | -8.82 | -10.16 | -10.15 | -11.51 | -11.06 | - |

Table 8: Perturbation energies $\Delta E$ of HCP clusters, reported in eV.

| Element | Diameter (Å) | $E_f$ | $E_0$ | $\Delta E_1$ | $\Delta E_2$ | $\Delta E_3$ | $\Delta E_4$ | $\Delta E_5$ | $\Delta E_6$ | $\Delta E_7$ | $\Delta E_9$ | $\Delta E_{11}$ |
|---|---|---|---|---|---|---|---|---|---|---|---|---|
| Tc | 2.735 | -11.52 | -114414.78 | -3.95 | -4.87 | - | -9.30 | -10.70 | - | -11.85 | -12.06 | -12.52 |
| Ru | 2.706 | -9.79 | -120855.98 | -4.66 | -5.07 | - | -8.67 | -9.78 | - | -10.84 | -11.10 | -11.37 |

d

## D  Cu-Bi grain boundary data

Grain boundaries are 2-dimensional defects that are present in real materials. They form the interface that separates the orientation in which atoms are stacked; and alter the microstructure, thus affecting performance. The data represented in graphical form in main text is provided in Table 9 and pictures of important clusters in Figure 5/

Table 9: Perturbation energies $\Delta E$ to the 14 atom first coordination shell surrounding the copper or bismuth atom at the center in the Σ5 grain boundary clusters Row 1: Number of atoms in the cluster. Row 2: Cluster radii in units of Å. Row 3: per atom energies of central cluster with Bi at center in eV – relative to isolated central cluster. Row 4: per atom energies of central cluster with Cu at center in eV – relative to isolated central cluster.

| Number of atoms in cluster | Cluster radii (Å) | $\Delta E$ with Bi (eV/atom) | $\Delta E$ with Cu (eV/atom) |
|---|---|---|---|
| 15 | 2.906 | 0.00 | 0.00 |
| 23 | 4.280 | -2.89 | 0.14 |
| 31 | 4.638 | -5.12 | -0.39 |
| 39 | 4.683 | -6.55 | -0.55 |
| 47 | 4.812 | -7.61 | -0.54 |
| 55 | 5.444 | -8.66 | -0.54 |
| 63 | 5.725 | -9.16 | -0.40 |
| 71 | 5.764 | -9.32 | -0.10 |
| 77 | 5.893 | -9.22 | 0.28 |
| 83 | 6.528 | -9.39 | 0.38 |
| 93 | 6.817 | -9.70 | 0.41 |
| 97 | 6.873 | -9.67 | 0.44 |

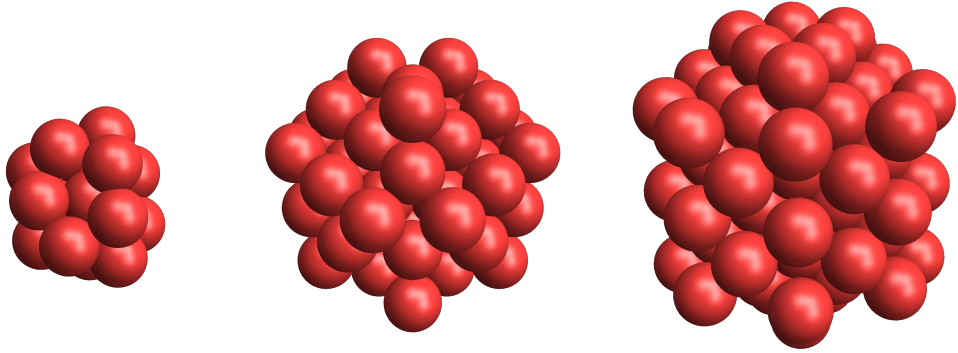

Figure 5: The grain boundary central 15-atom cluster (left), critical cluster consisting of 63 atoms (middle), and the largest cluster consisting of 97 atoms (right).

e

f



\end{document}